\documentclass[preprint]{aastex}
\usepackage{psfig}

\newcommand{\kms}{\, {\rm km\, s}^{-1}}
\newcommand{\erg}{\, {\rm erg}}

\newcommand{\cm}{\, {\rm cm}}
\newcommand{\seg}{\, {\rm s}}
\newcommand{\hz}{\, {\rm Hz}}
\newcommand{\ster}{\, {\rm sr}}
\newcommand{\mpc}{\, {\rm Mpc}}
\newcommand{\kpc}{\, {\rm kpc}}

\newcommand{\msun}{M_{\odot}}
\newcommand{\lya}{Ly$\alpha$ }
\newcommand{\hi}{\mbox{H\,{\scriptsize I}\ }}
\newcommand{\hei}{\mbox{He\,{\scriptsize I}\ }}
\newcommand{\heii}{\mbox{He\,{\scriptsize II}\ }}
\newcommand{\mgii}{\mbox{Mg\,{\scriptsize II}\ }}
\newcommand{\nhi}{N_{HI}}

\slugcomment{Submitted to ApJL}

\shorttitle{Self-shielding Effects}

\shortauthors{Zheng \& Miralda-Escud\'e}

\begin{document}

\title{Self-shielding Effects on the Column Density Distribution of
Damped Lyman Alpha Systems}

\author{Zheng Zheng \& 
Jordi Miralda-Escud\'e }
\affil{Department of Astronomy, The Ohio State University, 
Columbus, OH 43210}
\email{zhengz@astronomy.ohio-state.edu, jordi@astronomy.ohio-state.edu}

\begin{abstract}
  We calculate the column density distribution of damped \lya systems,
modeled as spherical isothermal gaseous halos ionized by the external
cosmic background. The effects of self-shielding introduce a hump in 
this distribution, at a column density 
$\nhi \sim 1.6\times 10^{17} X^{-1} \cm^{-2}$, where $X$ is the
neutral fraction at the radius where self-shielding starts being
important. The most recent compilation of the column density
distribution, by Storrie-Lombardi \& Wolfe, shows marginal evidence for
the detection of this feature due to self-shielding, suggesting a
value $X\simeq 10^{-3}$. Assuming a photoionization rate
$\Gamma\simeq 10^{-12} \seg^{-1}$ from the external ionizing
background, the radius where self-shielding occurs is inferred to be
$\sim 3.8 \kpc$. If damped \lya systems consist of a clumpy medium, this
should be interpreted as the typical size of the gas clumps in the
region where they become self-shielding. Clumps of this size with
typical column densities $N_H\sim 3\times 10^{20} \cm^{-2}$ would be
gravitationally confined at the characteristic photoionization
temperature $\sim 10^4$ K if they do not contain dark matter. Since
this size is similar to the overall radius of damped \lya systems in
Cold Dark Matter models, where all halos are assumed to contain similar
gas clouds producing damped absorbers, this suggests that the gas in
damped absorbers is in fact not highly clumped.
\end{abstract}


\section{INTRODUCTION}

  Damped \lya systems (hereafter, DLAs) contain most of the atomic
hydrogen in the universe (e.g., \citealt{Wolfe95,Storrie00}).  
They are therefore fundamental objects to understand
galaxy formation: in the process of forming galaxies, the matter that is
originally in the low-density, ionized intergalactic medium must
invariably become first atomic, before it can form molecular clouds and
stars.

  The nature and geometrical shape of the DLAs is still
a subject of debate. The two alternative hypotheses that are being
discussed at present and confronted with the observations are as
follows: (a) They are rotating disks, which may or may not be forming
stars; (b) They are approximately spherical halos, where global
rotation contributes little to the support against gravity. The truth
may be a combination of these two possibilities: DLAs
might be described as flattened halos or thick disks, with a range of
the ratio of rotation velocity to velocity dispersion.

  A fundamental property of DLAs is that their associated
metal lines reveal the presence of multiple absorbers in most of the
systems, which are thought to arise from clumps of gas. The multiple
absorption lines appear over velocity intervals of $30$ to $300 \kms$.
The statistical properties of the multiple absorption lines have been
used to attempt to distinguish between the disk and halo models, but
both alternatives seem to be consistent with observations 
(\citealt{Prochaska97,Prochaska98,Haehnelt98,Haehnelt00,McDonald99}).
Although the observations reveal that there are
usually one or a few clumps intersected along a random line of sight, with
characteristic column densities of $\sim 10^{20} \cm^{-2}$, 
the size of the clumps is not well known. Constraints in a few systems
from double lines of sight in lensed QSO's, and from model calculations
of the ionization parameter, indicate sizes or structure in the range
of 20 pc to 1 kpc (\citealt{Rauch99,Lopez99}).
Models of structure formation predict that the overall size of DLAs is of the
order of 1 to 10 kpc (e.g., \citealt{Katz96,Gardner97,McDonald99}).
If the clump size is not much smaller than the
size of DLAs, then the absorption lines would actually be arising from
mild fluctuations in density and velocity in a turbulent medium; whereas
if the clump sizes are much smaller, they would be real separate
entities with a large overdensity relative to an interclump medium.

  This paper discusses the effects of self-shielding on the form of the
column density distribution of damped \lya systems, and the dependence
of this distribution on the size of the gas clumps. As the gas becomes
self-shielding against the external cosmic ionizing background, there
is a rapid transition from ionized to fully atomic gas, and this
causes a feature in the column density distribution. If $X$ is the
neutral fraction of the gas seen at a column density $\nhi = 1.6\times
10^{17} \cm^{-2}$ (where the optical depth to photons at the hydrogen
ionization edge is equal to one), a hump in the column density
distribution will occur at $\nhi\sim 1.6\times 10^{17} X^{-1}
\cm^{-2}$. Therefore, a measurement of $X$ from this feature in the
column density distribution, combined with an independent estimate of
the ionizing background intensity, can be used to infer the size of
the gas clumps.

  Observationally, compared with the expected number extropolated from
low column density distributions, an excess of absorption systems with 
$\nhi \gtrsim 2\times 10^{20} \cm^{-2}$ is found by \citet{Lanzetta91}.
The recent compilation of the column density distribution of known DLAs
by \citet{Storrie00} shows more clearly a hump around $\nhi\sim 2\times 
10^{20}\cm^{-2}$. Theoretically, based on an approximate calculation of the 
self-shielding effect, \citet{Murakami90} showed that a flat part appears
in the column density distribution. With a simplified consideration on
the ionizing flux transfer, \citet{Petitjean92} investigated models of
self-gravitating, photoionized, spherical gaseous cloud and explained the
flattening of the column density distribution at $\nhi \sim 2\times 
10^{20} \cm^{-2}$ as the effect of self-shielding. Similar calculations 
of the expected column density distribution due to self-shielding have 
been done by \citet{Corbelli01} for plane-parallel geometry. 

Here, we will consider a spherical geometry with a singular isothermal
profile to perform a self-shielding calculation, 
and focus on the application of inferring the size of the gas clumps in 
damped \lya systems.

\section{METHOD}

\subsection{Model for Gaseous Halos}

  We model a DLA as a spherical cloud of gas with a singular isothermal
profile, $\rho_g = A/r^2$, with a cutoff at the virial radius, where 
$\rho_g$ is the gas density. The model depends only on the constant 
$A$, which can be recast in terms of the
mass of an associated virialized dark matter halo. We will be presenting
results for dark matter halo masses of $10^9$, $10^{10}$, $10^{11}$, and
$10^{12} \msun$, assuming that 5\% of the mass of the halo is in the
form of gas. For the cosmological model with $\Omega=1$, $H_0=70
\kms\mpc^{-1}$, and at $z=3$, the virial radii corresponding to these
halo masses are $r_{\rm vir} = 5.2, 11.1, 23.9,$ and $51.5\kpc$, and the 
halo circular velocities are $V_c= 29, 62, 134,$ and  $289\kms$, 
respectively (e.g., \citealt{Padmanabhan93}). In Cold Dark Matter models 
of structure formation, the damped \lya systems should be located in 
halos over this range of masses (e.g., \citealt{Gardner97}).
The constant $A$ is simply determined by the condition that
5\% of the halo mass is equal to the total gas mass within radius
$r_{\rm vir}$. The model ignores clumpiness of the gas, but as discussed
later the results depend only on the assumed radius at a given density,
so the halo radius will essentially play the role of the radius of a
clump.

\subsection{The Background Spectrum And The Ionization Profiles}

  The other quantity that determines the profile of neutral hydrogen is
the intensity of the external ionizing background. We assume an
intensity of $I_{\nu} = 3\times 10^{-22}
\erg\cm^{-2}\seg^{-1}\hz^{-1}\ster^{-1}$, constant at all frequencies
between the $\hi$ Lyman limit, $\nu_L$, and the \heii Lyman limit,
$4\nu_L$. The intensity is set to zero at frequencies above $4\nu_L$.
This is a sufficiently good approximation for our purpose to the
background spectrum expected from QSOs, after absorption by the \lya
forest is taken into account (e.g., \citealt{Haardt96}). The reason
is that \heii becomes self-shielding at a larger radius in the gaseous
halo than \hi (owing to the faster \heii recombination rate and the
lower intensity of the cosmic background at the \heii Lyman limit,
compared to hydrogen; e.g., \citealt{Miralda90}), and
therefore the photons at frequency above $4\nu_L$ in
the cosmic background are absorbed before reaching the region where
\hi opacity is important. With this spectrum, our adopted value of the
intensity at the \hi Lyman limit corresponds to a \hi photoionization
rate $\Gamma = 1.25\times 10^{-12} \seg^{-1}$.

  In our calculation, we will set the total hydrogen number density to
$n_H = 0.76 \rho_g/m_{H}$ (where $m_H$ is the mass of the hydrogen atom), 
and the electron number density to $n_e = 0.82 (1-x_{HI}) \rho_g/m_H$, 
where $x_{HI} = n_{HI}/n_H$ is the hydrogen neutral fraction. The factors 
0.76 and 0.82 assume a helium abundance of $Y=0.24$ by weight, and that 
all the helium is in the form of $\hei$ and $\heii$ with the same neutral 
fraction as hydrogen. Other than these factors, the presence of helium 
is ignored in our calculation. This
is a good enough approximation owing to the small helium abundance by
number, plus the fact that every photoionization of \hei produces a
photon following recombination of \heii which ionizes hydrogen.

\subsection{Self-shielding Calculation}

  We compute the self-shielded \hi density profile using an iterative
procedure similar to the one described by \citet{Tajiri98}.
Photoionization equilibrium
demands that at each radius, we have
\begin{equation}
\label{eqn:eqn1}
x_{HI}(r)\, \int_{4\pi} d\Omega \int_{\nu_L}^\infty
\frac{I_{\nu,0}e^{-\tau_\nu}}{h\nu} a_\nu d\nu =
[1-x_{HI}(r)]^2\, {0.82\rho_g(r)\over m_H}\, \alpha_B(T) ~,
\end{equation}
where $x_{HI}(r)$ is the hydrogen neutral fraction at radius $r$, 
$I_{\nu,0}$ is the cosmic background intensity, $\tau_\nu$ is the
photoionization optical depth from radius $r$ to infinity along a given
direction, $a_\nu$ is the photoionization cross section of hydrogen, and
$\alpha_B(T)$ is the case B recombination coefficient at temperature $T$ 
(we use the on-the-spot approximation, see \citealt{Osterbrock89}). 
We fix the temperature to $2\times 10^4$ K, for which the recombination
coefficient is $\alpha_B=1.41\times 10^{-13} \cm^3\seg^{-1}$ 
(\citealt{Spitzer78,Verner96}).
One can compute self-consistently the gas temperature assuming thermal
equilibrium under photoionization conditions, but in practice the gas
is likely to be frequently shock-heated in any model of moving gas 
clumps in a halo. To numerically evaluate the integral in 
equation (1), we compute the integral over frequency at many
values of the optical depth and obtain a fit to this function. The
integral over solid angle is reduced to one dimension in the spherical
case, and is performed at each radius. We initiate $x_{HI}(r)$ by
assuming the system to be optically thin. At each iteration,
$\tau_\nu$ is recomputed for each radius and polar angle, and new values
of $x_{HI}$ are found. This iteration is repeated until $x_{HI}$ has
converged with a fractional error less than $10^{-5}$.

\section{RESULTS}

\begin{figure}[ht]
\centerline{\psfig{figure=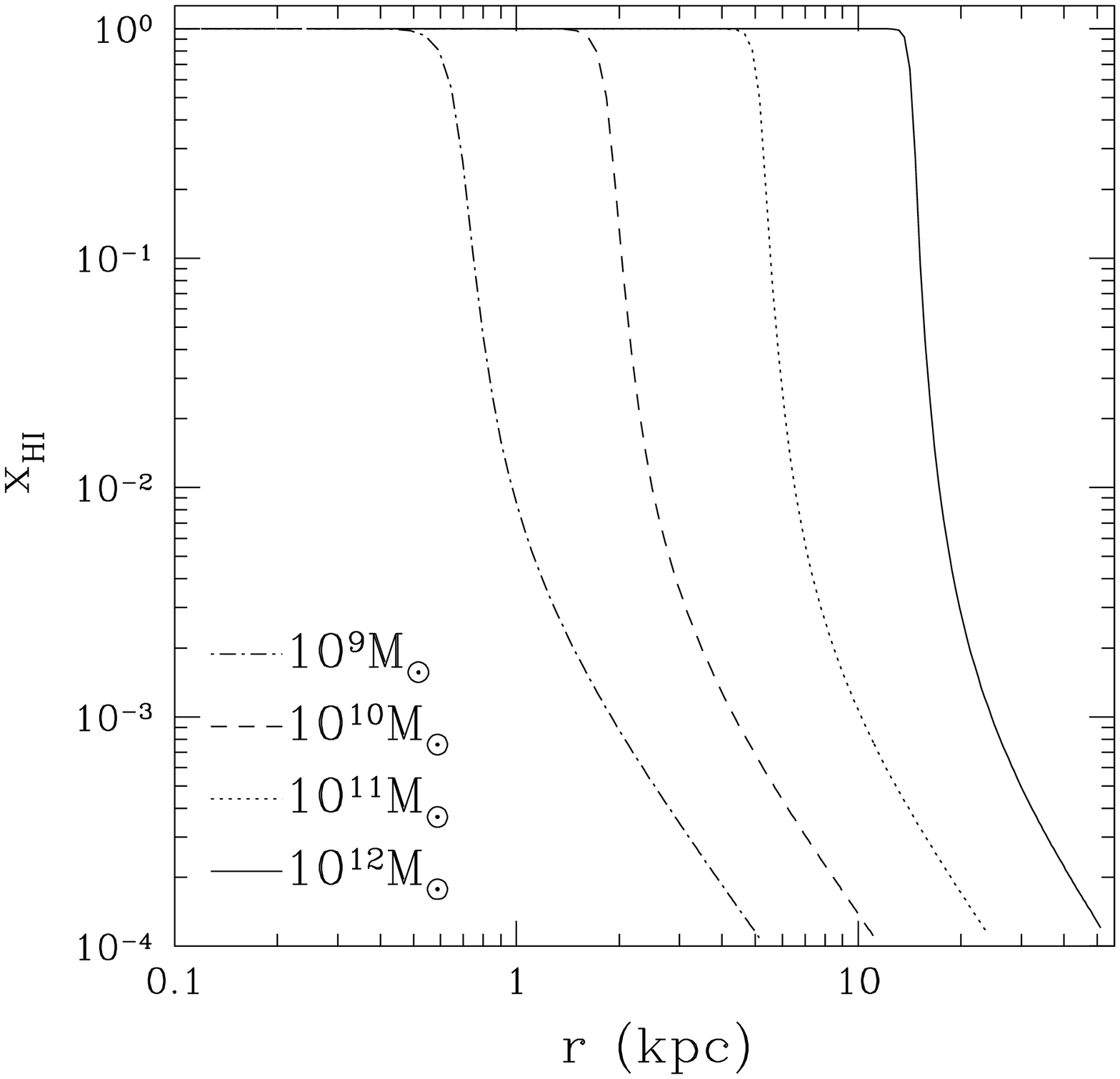,height=8cm,width=8cm} 
            \psfig{figure=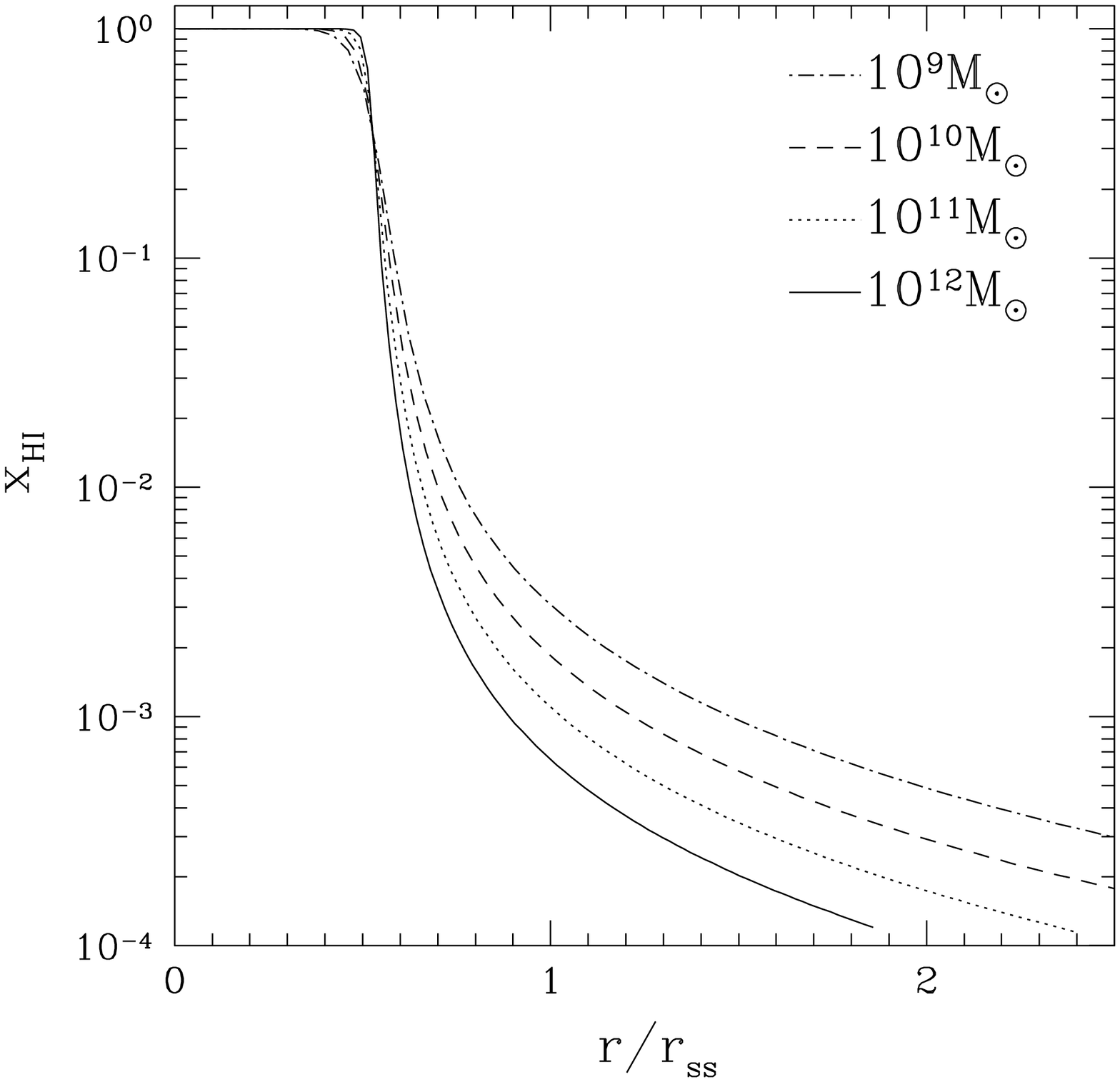,height=8cm,width=8cm} 
}

\caption[]{\label{fig:x_HI} The neutral fraction profile
as a function of radius for four halo masses. As explained in the text,
these $x_{HI}$ profiles depend only on the variable $X$, which is 
the neutral fraction at the self-shielding radius $r_{ss}$. 
The radius is in physical unit for the left panel and is normalized 
by $r_{ss}$ for the right panel. 
}
\end{figure}

  The neutral fraction profile as a function of radius, in $\kpc$, is
shown in the left panel of Figure 1 for the four halo masses. 
As expected, the ionizing photons penetrate deeper in halos of 
lower mass. For a power-law density profile, self-similarity implies 
that the shape of the $x_{HI}$ profile depends on the gas density 
and the background intensity only through one parameter, $X$, which is the
neutral fraction at the radius $r_{ss}$ where the cloud becomes
self-shielding: $X \equiv x_{HI,0}(r_{ss})$. Here, $r_{ss}$ is defined 
as the impact parameter where the column density is 
$N_{HI,0}(r_{ss}) = a_{\nu_L}^{-1} = 1.59\times 10^{17} \cm^{-2}$  
with both $x_{HI,0}$ and $N_{HI,0}$ computed without including the 
self-shielding correction. Using the
approximation $X \ll 1$ and ignoring also any external cutoff in the 
isothermal gas density profile, one finds 
\begin{equation}
 X = \left[ { 1.64 \alpha_B N_{HI,0} \over 0.76\pi\, \Gamma r_{ss} }
\right]^{1/2} ~.
\end{equation}
Thus, the shape of the neutral fraction profile depends only on the
product $\Gamma r_{ss}$. It is easily shown that the self-shielding
radius $r_{ss}$ is proportional to $(A^2/\Gamma)^{1/3} \propto
M^{4/9}/\Gamma^{1/3}$. For the four halo masses we use,
($10^9, 10^{10}, 10^{11},$ and $10^{12} \msun$), and 
$\Gamma = 1.25\times 10^{-12} \seg^{-1}$, we find
$r_{ss} = 1.3, 3.6, 9.9,$ and $27.6\kpc$, and $X= 1.8 \times 10^{-3},
1.1 \times 10^{-3}, 6.3 \times 10^{-4},$ and $3.8 \times 10^{-4}$, 
respectively. The right panel of Figure 1 shows the neutral fraction 
profiles for these four cases as a function of $r/r_{ss}$.

  This general shape of the neutral fraction profiles of self-shielding
clouds ionized by the cosmic background can be simply understood as
follows: the cloud becomes self-shielding at the radius where the total
rate of recombinations balances the rate at which photons come into the
cloud. Therefore, once a radius $\sim r_{ss}$ is reached where most
photons with frequency near $\nu_L$ have been absorbed, the cloud
becomes mostly neutral over a small range in radius, because the rate
of recombinations required to balance the absorption of all the higher
frequency photons is not much higher. The effect has been demonstrated
before by \citet{Murakami90} and \citet{Petitjean92}, and is analogous
to the sharp edges of \hi disks produced by self-shielding
\citep{Corbelli93}.  Since $a_{4\nu_L}\simeq a_{\nu_L}/64$, we conclude
that for $X> 1/64$, some of the highest frequency photons will penetrate
into the atomic region of the cloud (causing a residual ionization
inside the atomic region), while for $X < 1/64$, all the photons have
mean free paths small compared to $r_{ss}$ once the cloud is mostly
atomic, so the intensity of ionizing photons suddenly drops to zero and
there is a sharp transition to a completely atomic gas.

  Figure 2 shows the predicted distribution of \hi column densities,
$f(\nhi)$, when random lines of sight intersect the four model gas
clouds (obtained by requiring $f(\nhi)\, d\nhi \propto r\, dr$, where
$\nhi(r)$ is found by integrating the hydrogen density times the neutral
fraction in Fig. 1 over lines of sight at impact parameter $r$). The
transition to atomic gas causes a feature in the distribution, with a
hump appearing at column density $\nhi \sim N_{HI,0}\, X^{-1}$. 

\begin{figure}[h]
\centerline{\psfig{figure=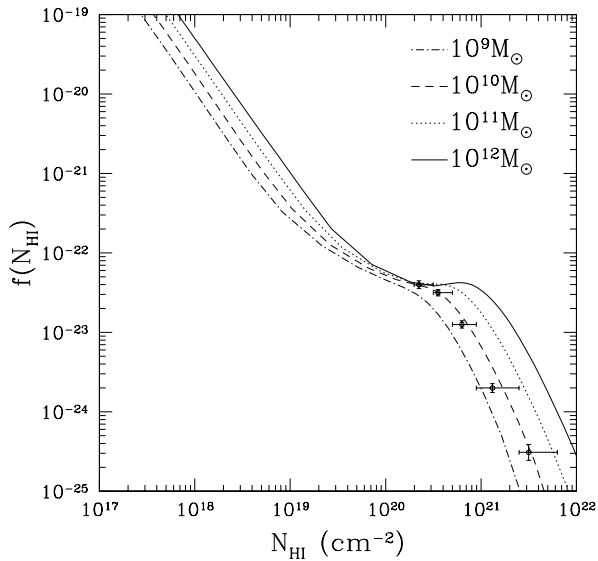,height=8cm,width=8cm} }
\caption[]{\label{fig:prob} The probability distribution of the neutral 
hydrogen column density for different halo masses. Each curve is 
normalized so that the total probability of intersecting the halo 
within the virial radius is unity. Data points are from \citet{Storrie00}
with a vertical shift.
}
\end{figure}

  Superposed on our curves, we reproduce a recent compilation of the
column density distribution of known damped \lya systems, from
\citet{Storrie00}. The points seem to fit best the curve 
for the $10^{10} \msun$ halo (notice that the points can be arbitrarily 
shifted vertically). In particular, the point at the lowest column density
suggests a shallower slope of the distribution. This is a marginal 
indication that the feature predicted by the self-shielding effect may 
have been detected. If so, the value of $X$ that is inferred is 
$X\simeq 1.1 \times 10^{-3}$, which for our adopted value of the 
photoionization rate $\Gamma$, corresponds to $r_{ss} \simeq 3.6 \kpc$.
We caution, however, that the predicted shape depends on the assumed
total gas density profile, which we have taken to be singular isothermal.

\section{DISCUSSION}

  We have shown that self-shielding should cause a sharp transition
from an ionized to an atomic medium in any gas cloud photoionized by
the external cosmic background. The expected absence of photons above
frequency $4\nu_L$ due to \heii absorption makes that transition even
sharper once the neutral fraction has been increased by a factor 64
by self-shielding. This produces a hump in the column density
distribution. The column density at which the hump occurs measures the
quantity $\Gamma r_{ss}$.

  Most damped \lya systems are likely to be photoionized by the external
cosmic background. In fact, since their rate of occurrence is about one
fourth per unit redshift at $z\sim 3$ \citep{Storrie96}, and
the mean free path of ionizing photons is $\Delta z \simeq 1$
\citep{Miralda90}, it is easily shown that if all the
sources of the cosmic ionizing background were embedded inside damped
\lya systems, then the local source in any such system would contribute
about the same flux as the external background on average. It is much
more likely that any sources associated with typical damped \lya systems
do not contribute significantly to the cosmic background.

  The results for the column density distribution presented by
\citet{Storrie00} suggest a possible detection of the
self-shielding effect, which would imply a radius 
$r_{ss} \simeq 3.6 \kpc$ for $\Gamma = 1.25\times 10^{-12} \seg^{-1}$. 
This represents a measurement of the size of the clumps
in damped \lya systems. Even though our calculation is for a spherical,
isolated cloud, a system of randomly located clumps would produce a
similar column density distribution depending on the radius of the
clumps, as long as each clump is illuminated by the ionizing background
with an optical depth $\lesssim 1$ along a large fraction of directions.
This should be correct since the typical number of intersected clumps
identified in the metal absorption lines is not very large.

  The typical size of damped \lya systems inferred from CDM models where
all halos with $V_c \geq 40 \kms$ give rise to the absorption systems 
is 1 to 10 $\kpc$ (e.g., \citealt{Katz96,Gardner97,McDonald99}). 
Hydrodynamic cosmological simulations show that, at redshift $z\sim 2-4$, 
for halos more massive than $\sim 10^{11}\msun$, the projected distance 
of DLAs to the center of the nearest galaxy is around $10 \kpc$ 
\citep{Gardner01}. Since the clump size we infer is about the same,
this suggests that the multiple metal absorption lines do not arise from
highly overdense clumps, but from mild density fluctuations in gaseous 
halos. If there were highly overdense clumps, their small self-shielding 
radii would produce a column density distribution with a less pronounced
hump at lower column density than the curves in Figure 2. The appearance
of individual absorption lines in the spectra of \mgii and other metal
lines might be deceiving, and could be due to a continuous medium, where
the \mgii density is highly sensitive to the gas density due to
photoionization and self-shielding effects.

  Nevertheless, a number of caveats must be borne in mind in
interpreting the observed column density distribution as a measurement
of $\Gamma r_{ss}$. In reality, there should be a distribution of
clump sizes, and differences from a simple spherical geometry would
introduce additional variations, which would smooth the shape of the
distribution predicted in Figure 2. Furthermore, the density profile of
the gas cloud can be different from what we assume here.
A fully three-dimensional
calculation of self-shielding in a hydrodynamic simulation of galaxy
formation would be highly desirable for a more robust interpretation
of the observations of damped \lya systems.

  Finally, we mention that once the size of the clumps is known,
their column densities can be used to infer their gas mass, and the
temperature required to have a cloud in hydrostatic equilibrium (see 
\citealt{Corbelli01}).
Using a typical column density $N_H\sim 3\times 10^{20}\cm^{-2}$ for
a clump, and a radius $r_{ss} \simeq 3.6 \kpc$, a gas mass of 
$ M_g \simeq 1.6 \times 10^8 \msun$ 
is inferred for the self-shielded region. In the absence of dark matter,
the temperature in hydrostatic equilibrium is $T \simeq G M_g m_H / k r_{ss} 
\simeq 2.3 \times 10^4$K. Since this is near the expected temperature in
photoionization equilibrium, this would imply that the gas clumps in 
damped systems do not contain a lot of dark matter. However, this conclusion
is altered if the gas temperature is higher due to shock heating or turbulent 
motions are important.

\acknowledgments

 This work was supported in part by NSF grant NSF-0098515.

{}

\end{document}